\documentclass[letterpaper,11pt]{article}
\usepackage{geometry} 
\usepackage[utf8]{inputenc}
\usepackage[T1]{fontenc}
\usepackage{lmodern}

\usepackage[dvipsnames,table]{xcolor}
\usepackage{graphicx}
\graphicspath{{./}{figures/}}
\usepackage{hyperref} 
\hypersetup{colorlinks = true, 
            urlcolor=RoyalBlue,
            citecolor=black,
            }
\usepackage{url}
\usepackage{gensymb}
\usepackage{fancyhdr}
\usepackage{wrapfig}
\usepackage{sidecap}
\usepackage{natbib}  
\usepackage{amsmath}
\usepackage{amssymb}
\usepackage{tcolorbox, xspace}
\usepackage{longtable}
\usepackage{enumitem}
\usepackage{floatrow}
\newfloatcommand{capbtabbox}{table}[][\FBwidth]
\usepackage{lipsum}
\usepackage[displaymath, mathlines]{lineno}
\modulolinenumbers[2]
\usepackage{float}
\usepackage{pdfpages} 
\usepackage{todonotes}
\usepackage{sectsty}
\usepackage{authblk}

\sectionfont{\fontsize{12}{15}\selectfont}
\subsectionfont{\fontsize{12}{15}\selectfont}

\newcommand{\RST}{{\it Roman}\xspace}
\title{\vspace{-3em}The Effects of HLTDS Cadence at Fixed Depth}

\definecolor{lightgray}{gray}{0.9}
\date{\vspace{-5ex}}

\begin{document}

\noindent
\textbf{\large Roman CCS White Paper: Optimizing the HLTDS Cadence at Fixed Depth}

\vspace{2em}
\begin{center}
{\Large }
\end{center}

\vspace{1.5em}
\noindent
\textbf{Roman Core Community Survey:} High Latitude Time Domain Survey

\vspace{0.5em}
\noindent
\textbf{Scientific Categories:} stellar physics and stellar types; stellar populations and the interstellar medium; large scale structure of the universe

\vspace{0.5em}
\noindent
\textbf{Additional scientific keywords:} Supernovae, Cosmology, Dark energy

\vspace{1em}
\noindent
\textbf{Submitting Author: David Rubin, University of Hawai`i (drubin@hawaii.edu)}\\

\vspace{0.5em}
\noindent
\textbf{List of contributing authors:}\\
Ben Rose, Baylor University (Ben\_Rose@baylor.edu)\\
Rebekah Hounsell, University of Maryland Baltimore County/ NASA Goddard Space Flight Center (rebekah.a.hounsell@nasa.gov)\\
Masao Sako, University of Pennsylvania (masao@sas.upenn.edu)\\
Greg Aldering, Lawrence Berkeley National Lab (galdering@lbl.gov)\\
Dan Scolnic, Duke University (daniel.scolnic@duke.edu)\\
Saul Perlmutter, University of California, Berkeley (saul@lbl.gov)\\
\vspace{1em}
\noindent

\thispagestyle{empty}
\newpage
\setcounter{page}{1}

\begin{figure}
    \centering
    \includegraphics[width=0.9\textwidth]{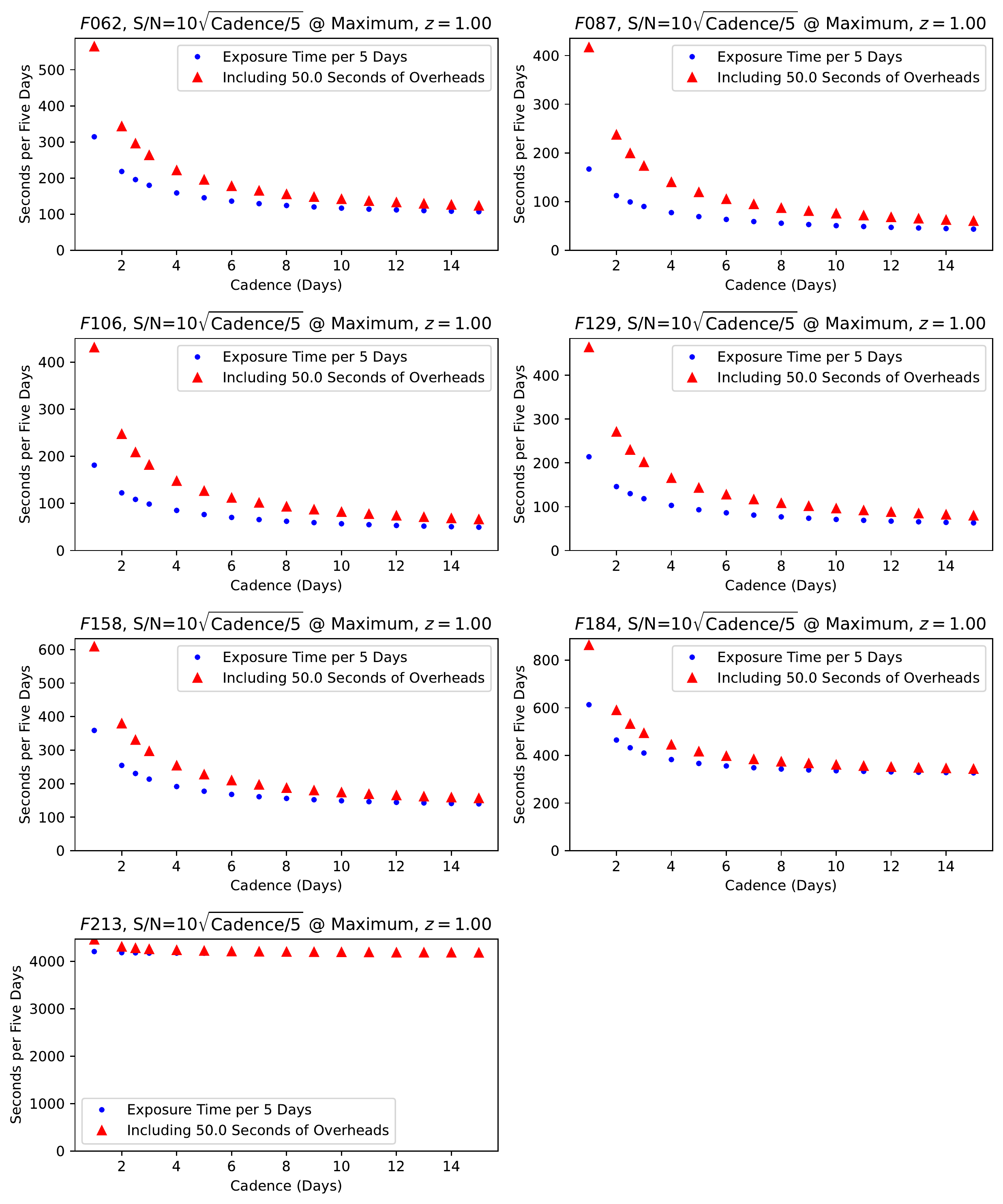}
    \caption{Exposure time (per 5 days) vs. cadence targeting S/N of  10$\sqrt{\text{cadence}}$ for the median SN~Ia at $z=1$. Each panel is computed for a different filter and shows both the exposure time itself (blue dots) and the total time assuming 50 seconds of overheads per visit (red triangles). The rapid decrease in exposure times with increasing cadence is due to the averaging down of read noise for longer exposures. The difference between the red triangles and blue dots is simply 50 seconds/(cadence/5 days).}
    \label{fig:ExpTimeOne}
\end{figure}

\begin{figure}
    \centering
    \includegraphics[width=0.9\textwidth]{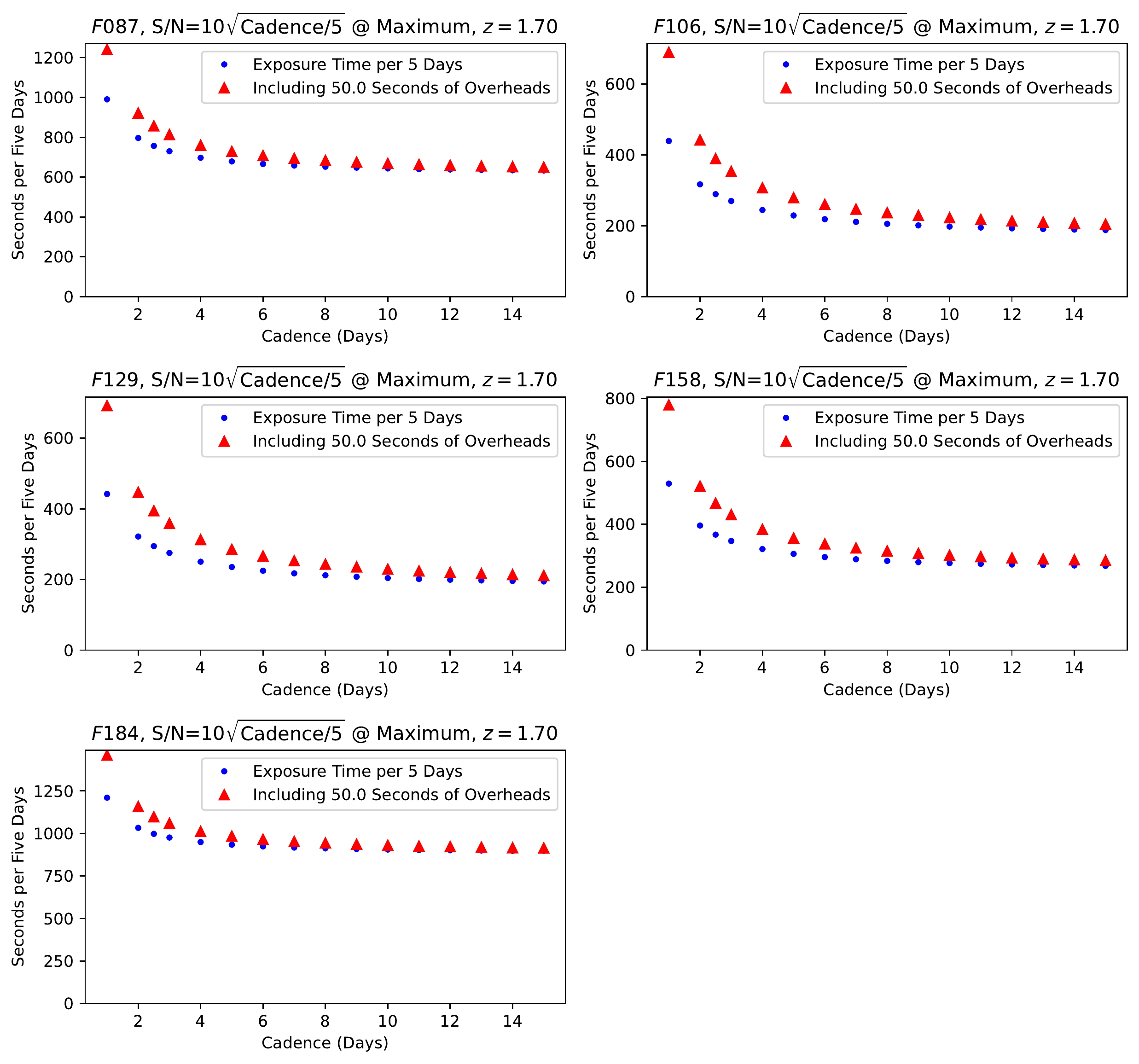}
    \caption{Exposure time (per 5 days) vs. cadence targeting S/N of 10$\sqrt{\text{cadence}}$ for the median SN~Ia at $z=1.7$. Each panel is computed for a different filter and shows both the exposure time itself (blue dots) and the total time assuming 50 seconds of overheads per visit (red triangles).}
    \label{fig:ExpTimeOneSeven}
\end{figure}

\begin{figure}
    \centering
    \includegraphics[width=0.95\textwidth]{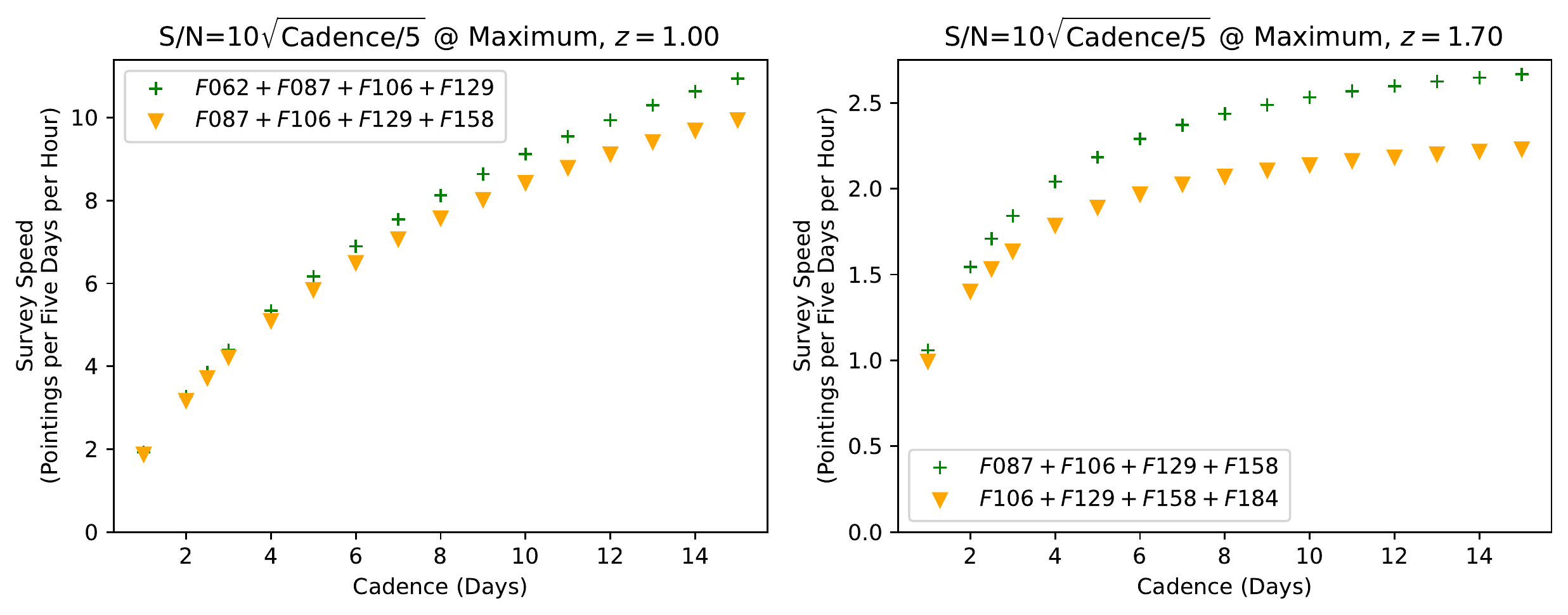}
    \caption{Summing over the red triangles in the previous figures (and dividing one hour into this total time), we find the total survey speed: the number of pointings per hour of survey time over five days. The left panel shows redshift 1 and the right panel shows redshift 1.7, both as a function of cadence. We show two different filter combinations for each redshift, finding similar results for both. The survey speed is much more sensitive to the cadence for the wide tier.}
    \label{fig:Rate}
\end{figure}


The current proposal for the High Latitude Time Domain Survey (HLTDS) is two tiers (wide and deep) of multi-band imaging and prism spectroscopy with a cadence of five days \citep{Rose2021c}. The five-day cadence is motivated by the desire to measure mid-redshift SNe where time dilation is modest as well as to better photometrically characterize the transients detected. This white paper does not provide a conclusion as to the best cadence for the HLTDS. Rather, it collects a set of considerations that should be used for a careful study of cadence by a future committee optimizing the \RST survey. This study should optimize the HLTDS for both SN~Ia cosmology and other transient science.

This white paper proposes that a fair way to trade cadence against the other survey parameters like survey area is to fix the survey depth per unit time, scaling the per-visit targeted S/N by $\sqrt{\text{cadence}}$. When fitting light curves and spectra using templates, fixing the survey depth per unit time in this way should give roughly the same SN parameter uncertainties (up to some maximum cadence to be determined). Figures~\ref{fig:ExpTimeOne} and \ref{fig:ExpTimeOneSeven} show our computed exposure times for the median SN~Ia at redshift 1 and redshift 1.7 (respectively) as a function of cadence from one day to fifteen days.

Longer required exposure times have more Poisson noise and are thus more amenable to faster cadence. The wide tier of the HLTDS has shorter exposure times and thus the survey speed is very sensitive to the cadence (Figure \ref{fig:Rate}, left panel). However, the cadence in the deep tier could be sped up (especially in the redder filters) without much loss in survey speed (Figure \ref{fig:Rate}, right panel). Thus the HLTDS has conflicting pressures put on it as SNe generally evolve faster at lower redshift and in bluer filters but these are conditions where longer cadence is more efficient.

The Committee should also consider the effects of staggering filters to improve cadence. For example, for a five-band survey, this could mean a single filter every day for five days, or a pair of filters (to provide a color) every two days.
    
The prism deep tier ($\sim 1$~hour every $\sim$~five days in the \citealt{Rose2021c} report) could approach a one-day cadence without much penalty to efficiency. In fact, prism survey strategies generally assume a four-point dither (e.g., \citealt{Rubin2022}), so these dithers could simply be on different days if desired.

Note that for the filter imaging (as opposed to the prism), there are currently no plans for chip-gap-spanning dithers (or any other kind of intra-filter dithers) during a single visit, as it is generally better to increase the cadence rather than dither. The fill factor of the \RST focal plane is about 7/8ths, so about 1/8th of the survey area will be missing in any one filter and the cadence will thus be twice the nominal value when this happens. Ideally, the survey should dither between different filters in the same cadence step to reduce the chances of missing two filters at the same time.

Fast-evolving transients always benefit from faster cadences (rather than better depth per visit at a slower cadence) for Euclidean geometry and Poisson-noise-dominated exposures. For example, for fast transients that only show up in one visit, increasing the cadence to one day from five days reduces the S/N per visit to 45\%, reducing the distance reach to 67\% and thus the sampled volume to 30\%. However, the number of visits increases by five, so the total number of detected fast transients in the survey increases to 150\%, in general scaling as (cadence)$^{1/4}$.

A point in favor of slower cadence is that detector noise has spatial/temporal correlations, so exposing long enough to average it down may improve systematic uncertainties in the photometry as well as improving statistical uncertainties. This should be evaluated using the observed noise from actual darks.

\bibliographystyle{apj}
\bibliography{library}

\end{document}